\documentclass[a4paper]{elsarticle}
\pdfoutput=1
\usepackage[colorlinks,bookmarks]{hyperref}
\usepackage[T1]{fontenc}
\usepackage[utf8]{inputenc}
\usepackage{lmodern}
\usepackage{amsmath}
\usepackage{enumerate}

\def\ppar{\nu}

\begin{document}

\begin{frontmatter}

\title{Coupling dark energy to dark matter inhomogeneities}

\author{Valerio Marra}
\address{Departamento de Física, Universidade Federal do Esp\'{\i}rito Santo,
29075-910, Vit\'oria, ES, Brazil}

\begin{abstract}
We propose that dark energy in the form of a scalar field could effectively couple to dark matter inhomogeneities. Through this coupling energy could be transferred to/from the scalar field, which could possibly enter an accelerated regime. Though phenomenological, this scenario is interesting as it provides a natural trigger for the onset of the acceleration of the universe, since dark energy starts driving the expansion of the universe when matter inhomogeneities become sufficiently strong. Here we study a possible realization of this idea by coupling dark energy to dark matter via the linear growth function of matter perturbations. The numerical results show that it is indeed possible to obtain a viable cosmology with the expected series of radiation, matter and dark-energy dominated eras. In particular, the current density of dark energy is given by the value of the coupling parameters rather than by very special initial conditions for the scalar field. In other words, this model -- unlike standard models of cosmic late acceleration -- does not suffer from the so-called ``coincidence problem'' and its related fine tuning of initial conditions.
\end{abstract}


\end{frontmatter}

\section{Introduction}
\label{sec:intro}

According to the standard model of cosmology -- confirmed by the latest results from ESA's Planck mission~\cite{Ade:2015xua} -- about 5\% of the energy content of the universe is made of ordinary baryons, that is, of particles belonging to the standard model of particle physics, recently glorified by the discovery of the Higgs boson~\cite{Aad:2012tfa,Chatrchyan:2012ufa}. The dark sector accounts for the remaining 95\%. More precisely, roughly 25\% consists of a yet-undetected matter component, which is thought to be a massive particle of non-baryonic nature that interacts through weak interaction and gravity only.
It is dubbed ``cold dark matter''.
Finally, dark energy is responsible for the missing 70\% of the energy content. The best candidate for dark energy to date is the so-called ``cosmological constant'', which is basically the energy of the vacuum and, in general relativity, is an arbitrary constant of nature. Its fundamental property -- gravitational repulsion -- causes the acceleration of the expansion of the universe~\cite{Perlmutter:1998np,Riess:1998cb}.

Faced with the formidable challenge of accounting for not only one but two unknown components, cosmologists studied at great depth dynamical models of dark energy~\cite{Wetterich:1987fm,Ratra:1987rm} and a possible interaction between dark energy and the other fields~\cite[see][and references therein]{Copeland:2006wr,2010deto.book.....A,Li:2011sd}, hoping to shed light on the nature of the dark sector.
In particular, a coupling between dark energy and dark matter~\cite{Wetterich:1994bg,Amendola:1999er,Dalal:2001dt,Zimdahl:2001ar} is suggested by the fact that the individual energy densities of the two components are today of the same order of magnitude. The latter is the so-called ``coincidence problem'', as an incredible fine tuning of the value of the cosmological constant is necessary for the standard model to explain observations.
However, a careful dynamical analysis of coupled dark energy models has shown that a fine tuning of the dark energy potential is unavoidable in order to have a viable cosmology~\citep[see, for example,][]{2010deto.book.....A}.

Here we propose that dark energy could effectively couple to dark matter inhomogeneities.
Indeed, if this is the case, the dark energy evolution will be altered --  possibly causing the universe acceleration -- at late time when inhomogeneities become strong. In other words, the trigger for acceleration is not given by tuned initial conditions but rather by the natural evolution of inhomogeneities.
We then take a phenomenological approach and consider a possible realization of this general idea by coupling dark energy to dark matter via the linear growth function of matter perturbations.
Our aim is to understand if such interaction can indeed give a viable dark energy dynamics free of fine tuning.

The outline of this paper is as follows. First we introduce the model in sections~\ref{sec:model}-\ref{sec:dysy}. Next we discuss the limiting cases in section~\ref{sec:limi} and the evolution in section~\ref{sec:evo}.
Conclusions are given in section~\ref{sec:conclu}.

\section{Model}
\label{sec:model}

Within General Relativity the total energy-momentum tensor is conserved and a possible interaction between dark matter and a scalar field can be modeled through an interaction current $Q^{\beta}$ which transfers energy and momentum from one source to the other:
\begin{equation} \label{inter}
\nabla_{\alpha} T_{m}^{\alpha \beta}= Q^{\beta}
\qquad
\nabla_{\alpha} T_{\phi}^{\alpha \beta}=- Q^{\beta} \,.
\end{equation}
In the individual conservation equations the interaction current has opposite sign so that the total energy-momentum tensor is conserved.

From equation~\eqref{inter} ($Q^{i}=0$ because of homogeneity and isotropy) one can derive the following system of dynamical field equations describing the background evolution of the well known Coupled Dark Energy model~\citep[see, e.g.,][]{2010deto.book.....A}:
\begin{align}
\label{friedmann}
& H^{2} = \frac{8 \pi G}{3} ( \rho_{m} + \rho_{\phi} +\rho_r ) \,, \\
\label{acce}
& \dot H + H^{2} = - \frac{4 \pi G}{3} ( \rho_{m} + \rho_{\phi} + 3 p_{\phi} + 2 \rho_r ) \,, \\
\label{continuityphi}
& \dot{\rho}_{\phi} + 3H (\rho_{\phi}+p_{\phi}) =  -Q^{0}  \,, \\
\label{continuitym}
& \dot{\rho}_{m} + 3H\rho_{m} =   Q^{0} \,, \\
\label{continuityrad}
& \dot \rho_r + 4H \rho_r =0 \,.
\end{align}
Curvature and the (subdominant) baryons are neglected, an overdot represents a derivative with respect to the cosmic time $t$, $H\equiv \dot{a}/a$ is the Hubble function, and $M_{\rm Pl}\equiv 1/\sqrt{8\pi G}$ is the reduced Planck mass with $G$ the Newton constant (we set $c=\hbar=1$).
The first two equations are, respectively, the Friedmann and the acceleration equation, while the last three equations are the continuity equations for the scalar field, matter and radiation, respectively.
The scalar field energy density, pressure and equation of state are, respectively:
\begin{align}
\rho_{\phi} = \frac{1}{2} \dot\phi^2 +V(\phi),
\quad
p_{\phi} =  \frac{1}{2} \dot\phi^2 -V(\phi),
\quad
w_{\phi}= \frac{p_{\phi}}{\rho_{\phi}} . 
\end{align}
%

\section{Interaction}
\label{sec:inter}

In the literature two classes of interactions -- with several variations -- have been studied:
\begin{align}
Q_{I}^{\beta} &= Q \, T_{m} \nabla^{\beta} \phi /M_{\rm Pl} \,,  \\
Q_{II}^{\beta} &= Q  \, T_{m} H \, u^{\beta}_{m} \,,
\end{align}
where $T_{m}=-\rho_{m}$ is the trace of the dark-matter energy-momentum tensor so that in equations \eqref{continuityphi} and \eqref{continuitym} one has:
\begin{align}
Q_{I}^{0}  &= Q \, \rho_{m} \, \dot \phi /M_{\rm Pl} \,,  \label{Q1} \\
Q_{II}^{0} &= Q  \, \rho_{m} \, H \,.
\end{align}
The first class follows from the interaction Lagrangian $\mathcal{L}_{\rm int}= m(\phi) \bar \psi \psi$ with $m(\phi)=m_{0} e^{Q \phi/M_{\rm Pl}}$,%
\footnote{Generally, the coupling is $Q=\partial \ln m(\phi) / \partial \phi.$}
that is, the mass of the dark matter field $\psi$ is a function of the scalar field~$\phi$, \citep[see][for details]{Wetterich:1994bg}.
The second class is instead an arbitrary parametrization, and equation~\eqref{inter} is seen as a phenomenological description for an effective interaction~\citep{Zimdahl:2001ar}.
As said earlier, careful dynamical analyses have shown that fine tuning of initial conditions is unavoidable in order to have a viable cosmology~\citep[see, for example,][]{2010deto.book.....A}.

Motivated by the fact that the evolution of inhomogeneities could trigger at the right time the onset of a dark-energy dominated universe, we now discuss the possibility that the dark-energy scalar field effectively couples to matter inhomogeneities.%
\footnote{The idea that the evolution of inhomogeneities could trigger the acceleration of the universe has been discussed in the literature regarding the so-called backreaction proposal, according to which late-time inhomogeneities could affect the average expansion rate of the universe, possibly explaining away dark energy (see, e.g., the special focus issue~\cite{Andersson:2011za}).}
The coupling is realized by considering an interaction like~\eqref{Q1}, in which the coupling is a function of the linear growth function of density perturbations $\delta_m$, $Q=\ppar \, \delta_m^n$, so that:%
\begin{equation} \label{Qd}
Q^{0} = \ppar \, \delta_m^n \, \rho_m \, \dot \phi/M_{\rm Pl} \,.
\end{equation}
%
The interaction \eqref{Qd} couples equations (\ref{continuityphi}-\ref{continuitym}) to the linear perturbation equation~\eqref{perteq}.
The use of $\delta_m$ is perhaps the simplest way to quantify the overall growth of inhomogeneities in the universe.
The dimensionless coupling parameter $\ppar$ sets the strength of this interaction.
The exponent $n$ parametrizes the dependence of the effective coupling parameter $Q=\ppar \, \delta_m^n$ on the linear growth function $\delta_m$.
For the sub-horizon scales at which a canonical scalar field remains homogeneous, the growth function does not depend on the wavenumber $k$ and can be obtained numerically from the standard equation~\cite{2010deto.book.....A}:
\begin{equation} \label{perteq}
\ddot \delta_m +2 H \dot \delta_m - 4 \pi G \rho_m \delta_m =0  \,.
\end{equation}
As usual, we normalize $\delta_m$ to unity at the present time, $\delta_m(t_0)=1$.
It is interesting to note that the coupling parameter $\ppar$ does not appear as the correction due to $\ppar$ is of order $n+1$ in $\delta_m$.
Therefore, the growth of perturbation is altered only by the different background evolution, meaning that this model could agree with growth data for arbitrarily large values of $\ppar$.

\subsection{Discussion}
\label{sec:disc}

Although in the following we take the phenomenological approach in which we assume the validity of \eqref{Qd} and examine its cosmological implications, it is interesting to discuss which mechanism could be behind such an interaction.

Physical processes that are linked to the growth of $\delta_{m}$ are necessarily nonlinear, such as collapse and shell crossing.
However, quantities associated to scalar perturbations do not seem suitable for the coupling~\eqref{Qd}. Indeed, local values of e.g.~matter density and velocity divergence can be arbitrary high in the past even if $\delta_{m}$ was initially very small.
The obvious quantity that at early times is absent (or negligible) is the vorticity (or curl) of the velocity field,%
\footnote{Not to be confused with the halo spin, see~e.g.~\cite{Libeskind:2012ya}.}
which is generated at very non-linear stages of the dark matter collapse. More precisely, it is produced when, at shell crossing, the single-stream irrotational description of the dark matter fluid ceases to be valid~\cite{Pichon:1999tk}. 
Therefore, as time progresses, vorticity is produced on ever larger scales.
It is difficult to study analytically the production of vorticity and many works have used dedicated $N$-body simulations in order to obtain its magnitude and time/scale dependence~\cite{Wang:2013gtn,Libeskind:2013zea,Hahn:2014lca}.
In particular, the results of~\cite{Libeskind:2013zea,Hahn:2014lca} show how vorticity is correlated with the density contrast and that it becomes important (as compared to the velocity divergence) at small scales and/or at early times.

The coupling of~\eqref{Qd} could then arise if the scalar field is coupled to the vector perturbations associated with vorticity. Phenomenologically, the coupling $Q$ will be proportional to an invariant contraction of the gradient tensor of the velocity field $\partial v_{i}/ \partial x_{j}$ (see~\cite{Wang:2013gtn} for a discussion of possible invariants), which, based on the results from numerical simulations~\cite{Libeskind:2013zea,Hahn:2014lca}, we can parametrize through a power of $\delta_{m}$, as in the expression $Q=\ppar \, \delta_m^n$ above. Note, however, that we are parametrizing the nonlinear growth of perturbations (the one correlated with the vorticity) with the linear growth function of equation~\eqref{perteq}.
Although this is similar in spirit to what is usually done within the spherical top-hat model for nonlinear collapse, it is also clear that this approach will only be able to describe qualitatively the growth of vorticity.

The growth function $\delta_{m}$ of equation~\eqref{perteq} describes the normalized growth of perturbations. Clearly, on smaller scales perturbations are stronger (i.e.~the initial value of $\delta_{m}$ is larger) and nonlinearity is reached at earlier times, as already mentioned in the discussion above. It is then natural to question whether a scale independent quantity such as $\delta_{m}$ is able to parametrize the growth of vorticity.
From the results of~\cite[][Fig.~3]{Libeskind:2013zea} one sees that the magnitude $\omega(r,z)$ of the vorticity grows at approximately the same rate at different scales.
Or, equivalently, at the various redshifts examined the scale-dependence of $\omega(r,z)$ follows a power law with approximately constant slope.
Therefore, if one were to average the vorticity contribution at fixed times, one would obtain an approximately constant factor times the vorticity at some pivot scale $r_{p}$: $\langle \omega(r,z) \rangle \approx \text{const} \, \omega(r_{p},z)$.
The constant factor can then be absorbed into the coupling constant $\ppar$ and, as discussed above, $\omega(r_{p},z)$ can be parametrized using $\delta_{m}$ so that one obtains $Q=\ppar \, \delta_m^n$.
This should justify the approximation of using the scale-independent quantity $\delta_{m}$ in order to parametrize the scale-dependent onset of vorticity.

Although we have justified -- at least in a first approximation -- the use of the linear growth function $\delta_{m}$ as a proxy for the vorticity of the velocity field, it would be desirable to quantify directly and more precisely the growth of vorticity using numerical simulations.

\section{Dynamical system approach}
\label{sec:dysy}

As customary~\cite{Copeland:1997et,Amendola:1999er}, let us switch to the following variables:
\begin{align}
x^{2} &=  \frac{\dot{\phi}^{2}}{6M_{{\rm Pl}}^{2}H^{2}} ,
\qquad
y^{2} = \frac{V}{3M_{{\rm Pl}}^{2}H^{2}} , \nonumber \\
\Omega_m &= \frac{\rho_m}{3M_{{\rm Pl}}^{2}H^{2}},
\qquad
\Omega_r = \frac{\rho_r}{3M_{{\rm Pl}}^{2}H^{2}} \,.
\end{align}
The Friedmann equation \eqref{friedmann} becomes then:
\begin{equation} \label{fried2}
1=x^2+y^2 +\Omega_m+ \Omega_r \,,
\end{equation}
so that one variable is not independent.
The dark energy density parameter and equation
of state, and the effective equation of state are, respectively:
\begin{align}
\Omega_{\phi}= x^2 + y^2,
\quad
w_{\phi}  =  \frac{x^{2}-y^{2}}{x^{2}+y^{2}},
\quad
w_{{\rm eff}}  =  x^{2}-y^{2}+ \frac{\Omega_r}{3} .
\end{align}

We then consider the case of an exponential self-interaction potential such as:
\begin{equation}
V(\phi)= V_0 e^{- \lambda \phi/M_{\rm Pl}} \,,
\end{equation}
where $\lambda$ is the dimensionless self-interaction coupling.
Using the e-folding time variable $N\equiv\ln (a/a_0) = - \ln (1+z)$ (we will denote the derivative with respect to $N$ with a prime) equations (\ref{acce}-\ref{perteq}) become:
\begin{align}
& \frac{H'}{H}  +  \frac{3}{2}\left(1+ x^2-y^2 +\frac{1}{3} \Omega_r \right) =0 \,,\label{eq:Hprime}\\
& x' + \left( \frac{H'}{H} +3 \right)x - \frac{\sqrt{6}}{2} \lambda \, y^2 =-\frac{\sqrt{6}}{2}    \ppar\,   \delta_m^n \, \Omega_m\,, \label{eq:contp} \\
& y' + \left(\frac{H'}{H}  + \frac{\sqrt{6}}{2} \lambda \, x \right) y =0  \,, \\
& \Omega_m'  + \left (2 \frac{H'}{H} +3 \right) \Omega_m  = + \sqrt{6} \,  \ppar\,  \delta_m^n \, \Omega_m \, x\,, \label{eq:contm} \\
& \Omega_r' + \left(2 \frac{H'}{H} +4 \right) \Omega_r =0 \,, \label{eq:contr} \\
&  \delta_m'' + \left(  \frac{H'}{H} +2  \right)  \delta_m' - \frac{3}{2}\Omega_m   \delta_m =0  \,. \label{eq:pertN}
\end{align}
%

\section{Limiting cases}
\label{sec:limi}

We will now discuss the known limiting cases of this model.
First, for $\lambda=\ppar =0$ one recovers the $\Lambda$CDM model, with $\Omega_{\Lambda 0}=y(0)= V_0/\rho_{c0}$.

For $\ppar=0$ the system of equations (\ref{eq:Hprime}-\ref{eq:pertN}) reduces to the standard case of an uncoupled quintessence field with exponential potential.
In particular, if $\lambda^2<3$ there is a stable (accelerated for $\lambda^2<2$) attractor with $\Omega_\phi=1$ and $w_{\phi}=w_{\rm eff}=-1+ \frac{\lambda^2}{3}$ (scalar field-dominated solution); and, if $\lambda^2>3$, a stable attractor with $\Omega_\phi=\frac{3}{\lambda^2}$ and $w_{\phi}=w_{\rm eff}=0$ (scaling solution).

For $n=0$ one recovers the case of coupled dark energy with $Q=\ppar$~\cite{Wetterich:1994bg,Amendola:1999er}.
In this case, if $\lambda^2<2$ there is an accelerated fixed point with $\Omega_\phi=1$ and $w_{\phi}=w_{\rm eff}=-1+ \frac{\lambda^2}{3}$.
This point is stable for $\lambda (Q+\lambda)<3$.
If instead it is $\lambda (Q+\lambda)>3$, one has a stable fixed point with $\Omega_\phi=\frac{Q(Q+\lambda)+3}{(Q+\lambda)^2}$, $w_{\phi}=\frac{-Q(Q+\lambda)}{Q(Q+\lambda)+3}$ and $w_{\rm eff}=\frac{-Q}{Q+\lambda}$. This fixed point is accelerated if $Q<- \lambda$ or $Q> \lambda/2$.
This scaling solution can in principle explain the acceleration of the universe without fine tuning. However, it does not produce a viable cosmology as the matter dominated era does not set in and matter inhomogeneities do not grow~\cite{2010deto.book.....A}.

For $n \neq 0$ -- the case considered here -- it turns out that exists only one fixed point, characterized by $\delta_m^n(\infty)  \Omega_m(\infty) = \lambda/\ppar $, for which it is possible to obtain also the asymptotic behavior of all the dynamical functions~\cite{paper2}.
However, as we will see, for a qualitative understanding it is very useful to define ``instantaneous'' fixed points using the instantaneous coupling $Q(N)=\ppar \, \delta_m^n(N)$.

\section{Evolution}
\label{sec:evo}

\begin{figure}
\begin{center}
\includegraphics[width=.49 \columnwidth]{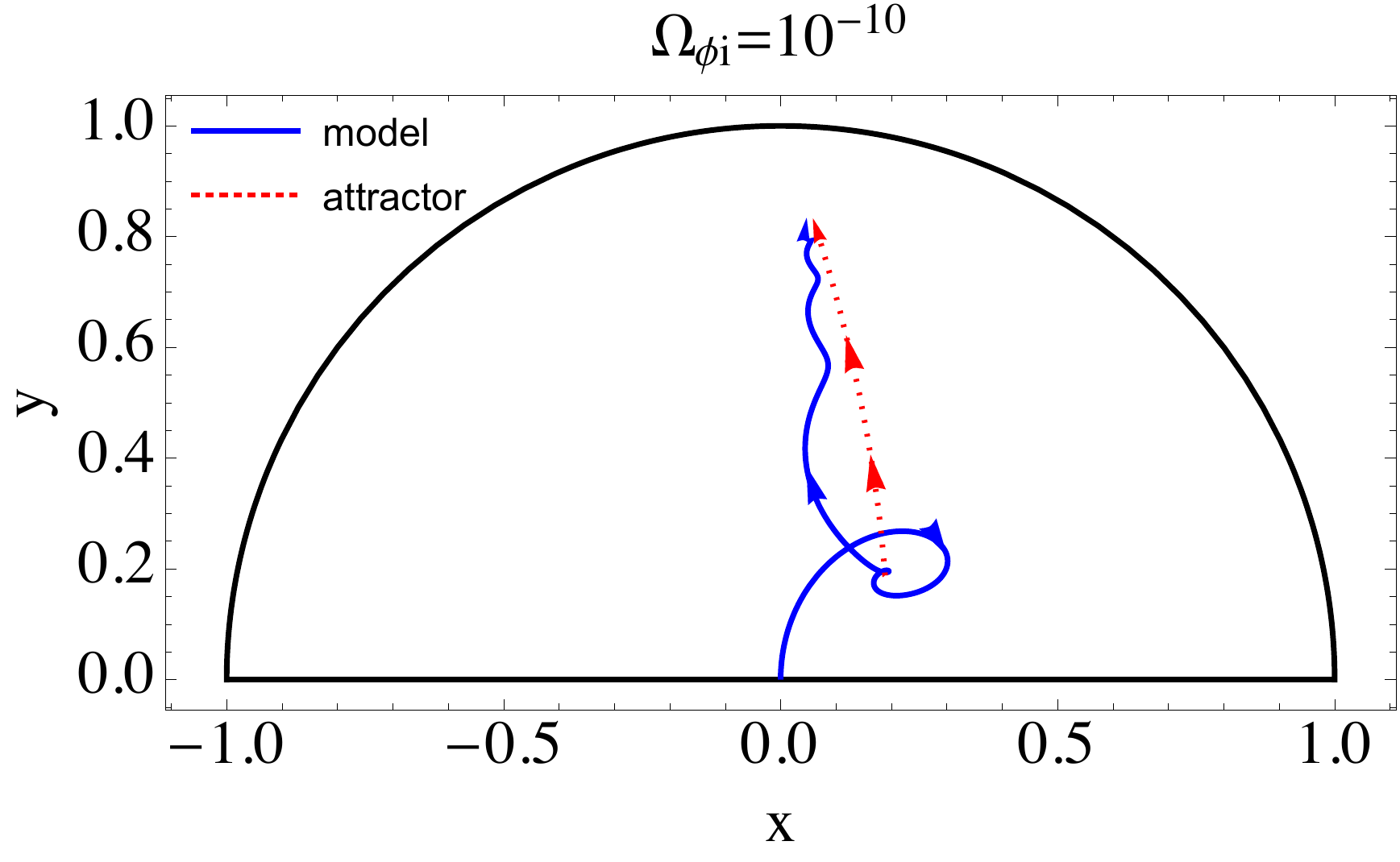}
\includegraphics[width=.49 \columnwidth]{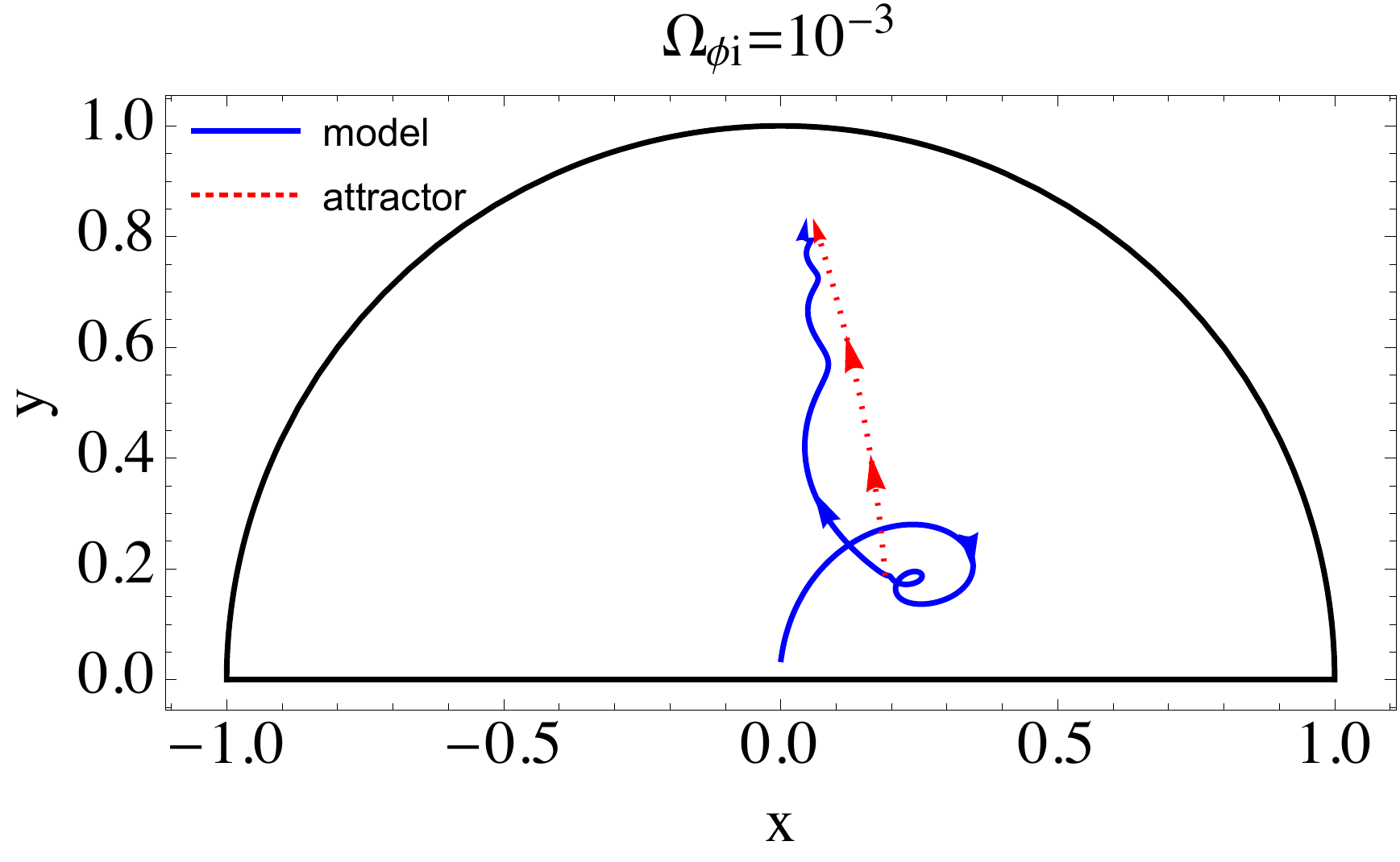}
\includegraphics[width= .49 \columnwidth]{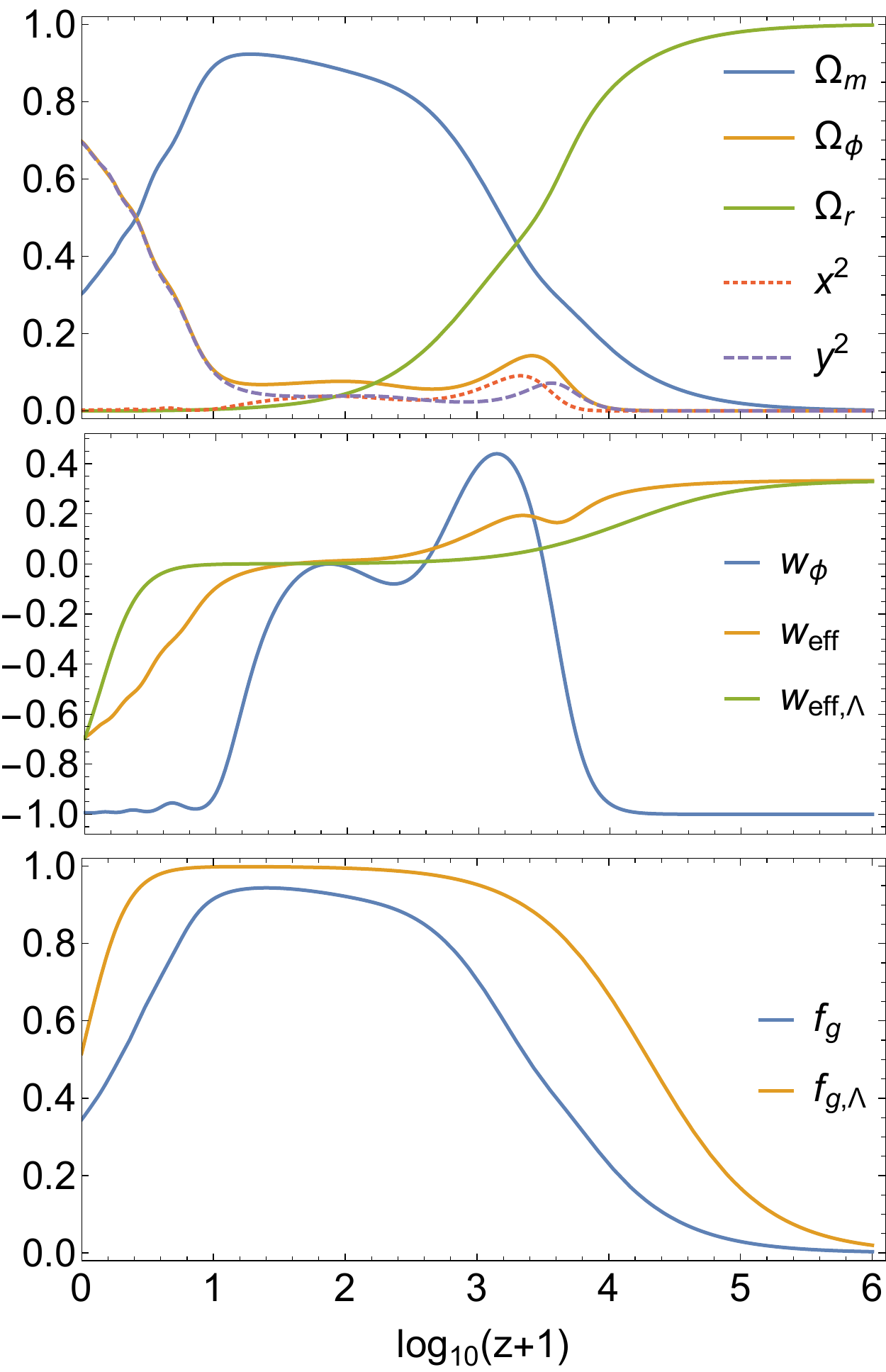}
\includegraphics[width= .49 \columnwidth]{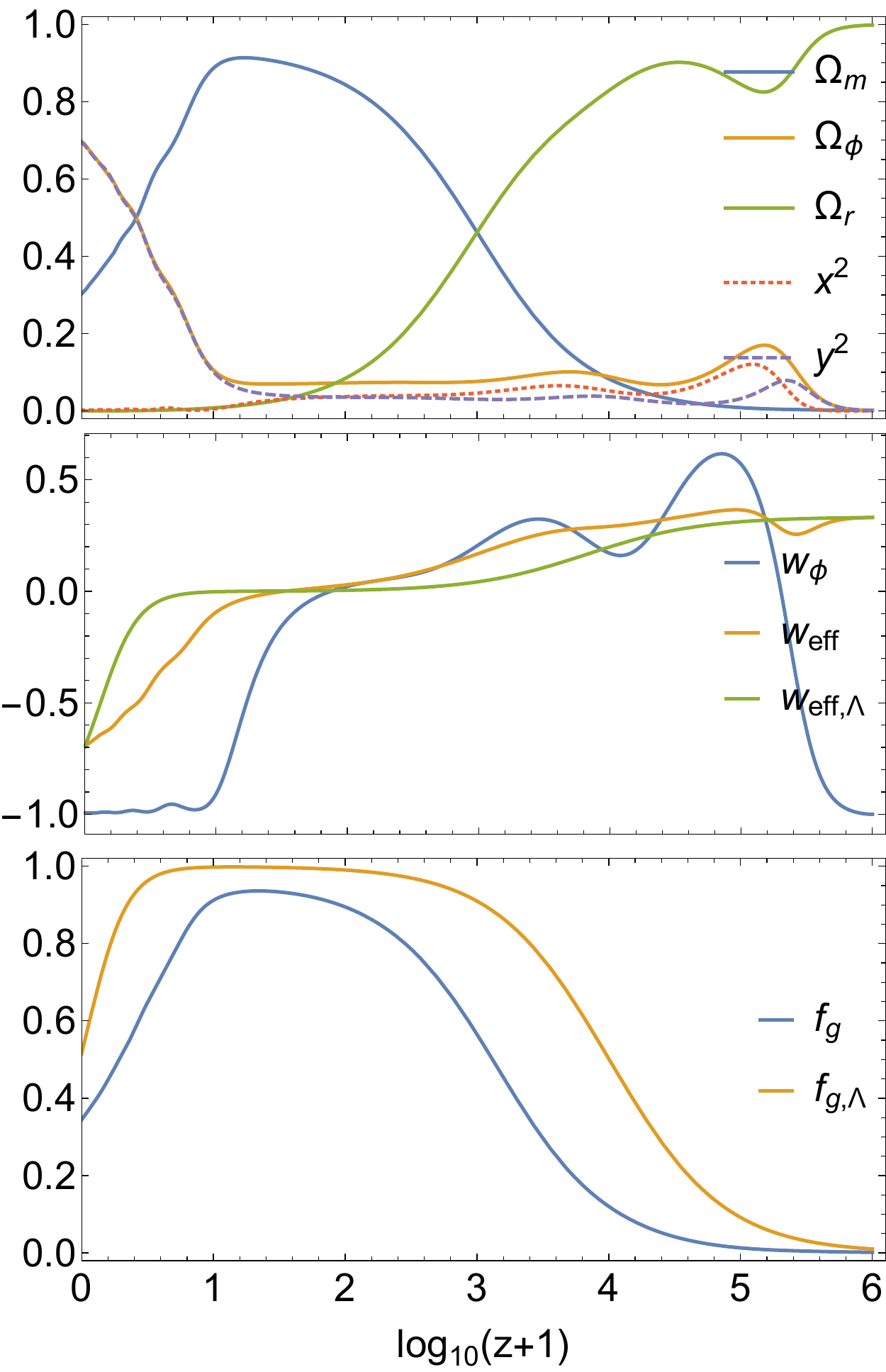}
\caption{
Comparison of the dynamical evolution of the model presented in this paper with parameters $\{ \lambda , \ppar, n \}= \{6.5 ,14.5, 2 \}$ for two very different sets of initial conditions.
The plots of the left column are relative to initial conditions $\{\Omega_{\phi i}, \Omega_{ri} \}= \{10^{-10}, 0.998 \}$ at the initial redshift $\log_{10}(1+z)=6$, while the plots of the right column are relative to $\{\Omega_{\phi i}, \Omega_{ri} \}= \{10^{-3}, 0.998 \}$ at the same initial redshift.
\emph{First row:}
trajectory in the $\{ x, y\}$ phase space of the solution of equations~(\ref{eq:Hprime}-\ref{eq:pertN}) (solid blue) and of the instantaneous attractor (dotted red, see section~\ref{sec:limi}).
The black curve is the border of the allowed region given by $y=\sqrt{1-x^2}$.
Note how the model converges to the present-day accelerated scaling instantaneous attractor. 
\emph{Second row:}
redshift evolution of the relevant density parameters. The dynamics features the expected series of radiation, matter and dark-energy dominated eras.
\emph{Third row:}
evolution of the dark energy equation of state and of the effective equation of state.
Also shown for comparison is the effective equation of state for a flat $\Lambda$CDM model with the same present-day values of the density parameters.
\emph{Fourth row:}
evolution of the growth rate $f_g= \delta_m'/ \delta_m$ compared to the growth rate relative to the corresponding flat $\Lambda$CDM model defined above. 
As can be seen by comparing the plots of the left column with those of the right column the model of this paper does not suffer from the fine tuning of initial conditions: the dynamics is basically unaltered if the initial dark energy density is changed by seven orders of magnitude.
}
\label{evo}
\end{center}
\end{figure}

The basic idea of the model of equations~(\ref{eq:Hprime}-\ref{eq:pertN}) is that a coupling to matter perturbations makes the scalar field enter a scaling accelerated solution. From the discussion above on the limiting cases one concludes that this is achieved if the scalar field reaches today the attractor with $\Omega_\phi=\frac{Q(Q+\lambda)+3}{(Q+\lambda)^2}$, where $Q=\ppar \, \delta_m^n(0) = \ppar$.
At early times, this coupling will be suppressed by the small early values of the growth function $\delta_m$.
The growth function provides therefore an effectively time-dependent coupling with a step-like functional form. This is exactly what is needed in order to have a matter dominated era followed by the scaling attractor above, as first advocated by Amendola and Tocchini-Valentini~\cite{Amendola:2000uh}, in which a transition to the accelerated attractor was obtained by means of a $\phi$-dependent coupling (see also~\cite{Baldi:2011wy}).
We will adopt as values for $\lambda$ and $\ppar$ the present-day values used in~\cite{Amendola:2000uh}.
Within our notation these values are $\{ \lambda , \ppar \}= \{6.5 ,14.5 \}$ and give a scaling accelerated attractor with $\Omega_{m0}\simeq 0.3$ and $\Omega_{\phi 0}\simeq 0.7$.
Note that, as both couplings are positive, energy is transferred from dark energy to dark matter.
The model parameters are now fixed (we will set $n=2$) and only the initial conditions are left to be chosen.

Figure~\ref{evo} (left column) shows the dynamics for the initial conditions $x_i=0$, $y_i=10^{-5}$ -- which correspond to  $\Omega_{\phi i}=10^{-10}$ -- and $\Omega_{ri}=0.998$ at the initial redshift $\log_{10}(1+z)=6$.
Figure~\ref{evo} (right column) shows instead the dynamics for $x_i=0$, $y_i=10^{-1.5}$ -- which correspond to  $\Omega_{\phi i}=10^{-3}$ -- and $\Omega_{ri}=0.998$ at the same initial redshift.
The initial conditions for equation~\eqref{eq:pertN} have to be given taking into account the early radiation dominated era. Specifically, they are
\begin{equation}
\tilde \delta_m(N_i) =e^{N_i}+ \frac{2}{3} e^{N_{\rm eq}} ,
\qquad
\tilde \delta_m'(N_i) = e^{N_i} ,
\end{equation}
where $N_{\rm eq}$ is the e-folding time at which $\Omega_r(N_{\rm eq})=\Omega_m(N_{\rm eq})$~\citep[for details, see][]{1980lssu.book.....P}.

In figure~\ref{evo} (first row, left panel), the model trajectory (solid blue) in the $\{ x, y\}$ phase space reaches the point $\{ x_0, y_0\}=\{ 0.0468, 0.833 \}$, the instantaneous attractor trajectory (dotted red) reaches instead the point $\{ 0.0583, 0.833 \}$. The field is basically on the instantaneous attractor, which is then useful for understanding the dynamics.
Also, it is clear from the plot how the initial attractor was frozen (and not accelerated) for a long time allowing the model trajectory to spiral around it. Then the effective coupling started making the attractor move towards its present location.

\begin{figure}
\begin{center}
\includegraphics[width= .6 \columnwidth]{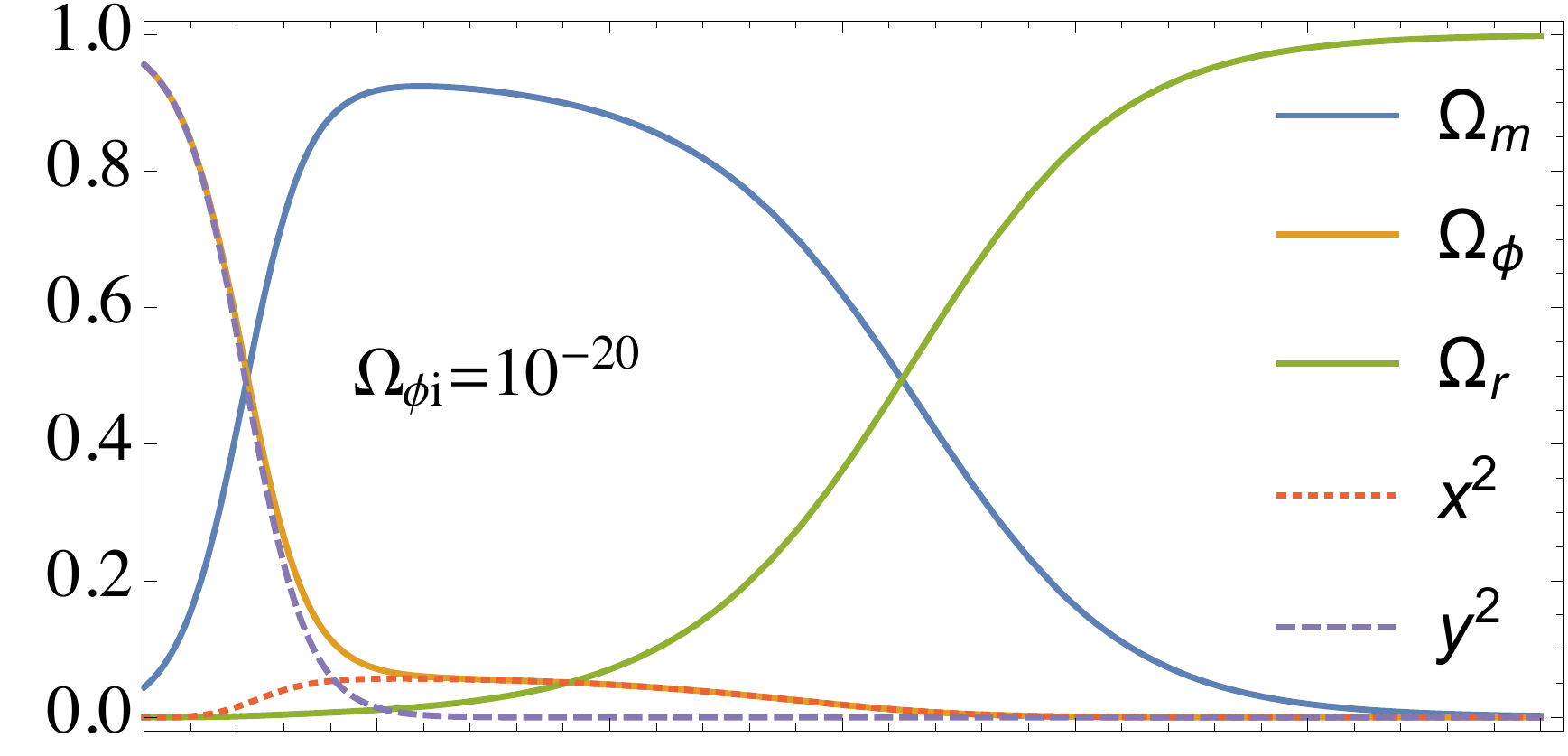}
\includegraphics[width= .6 \columnwidth]{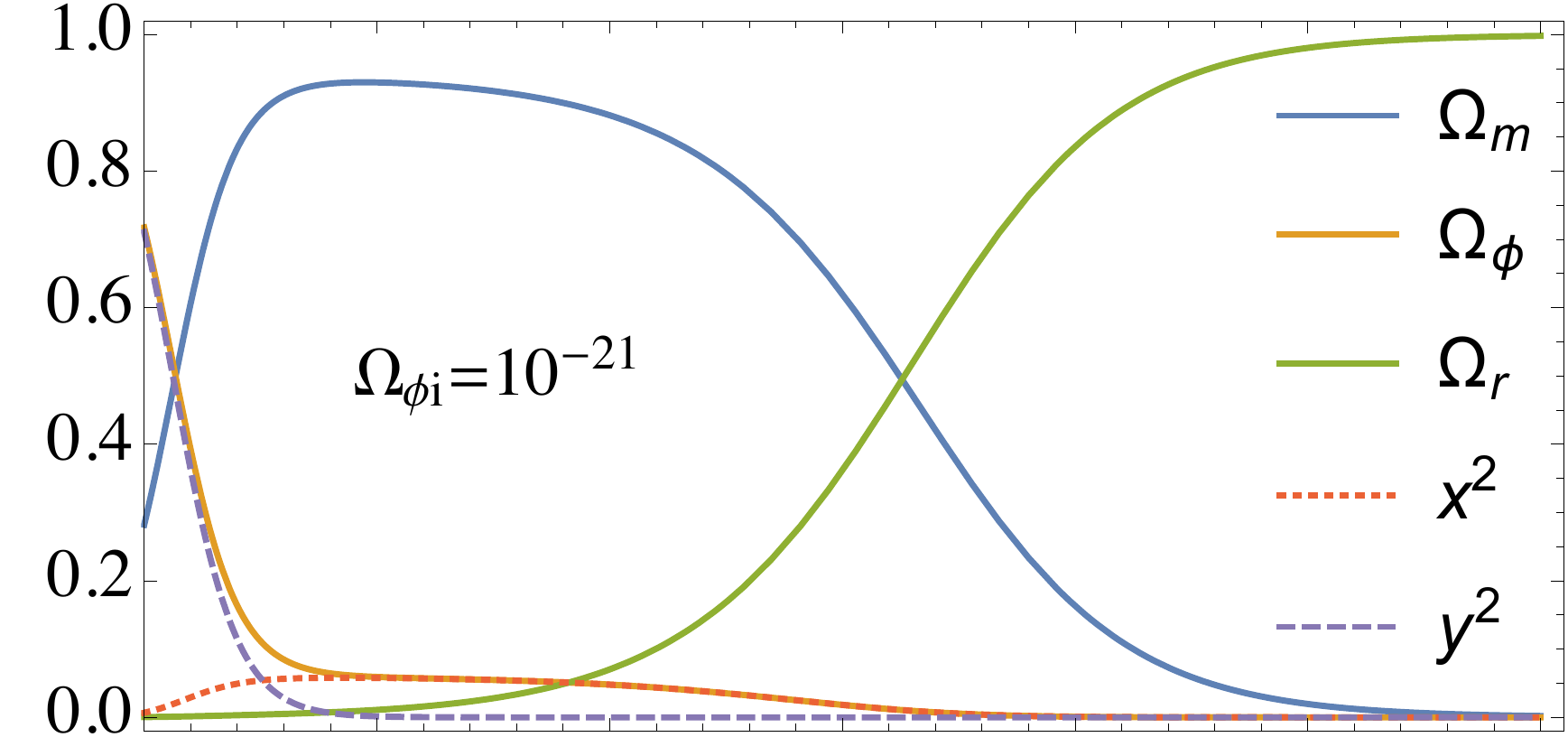}
\includegraphics[width= .6 \columnwidth]{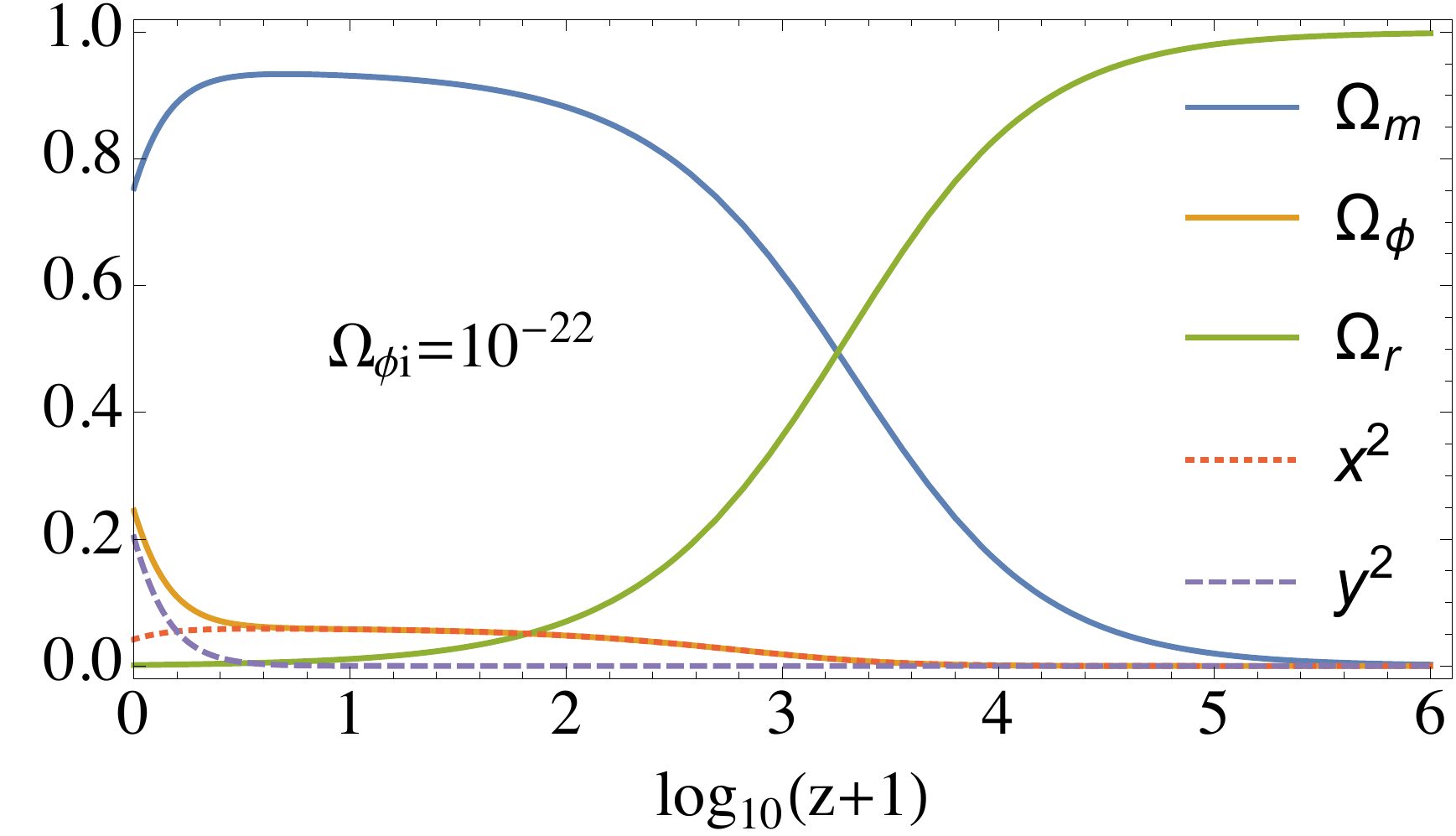}
\caption{
Redshift evolution of the density parameters for the standard coupled dark energy model ($n=0$ within the notation of this paper) with $\{ \lambda , \ppar \}= \{0.1 ,0.3 \}$.
The accelerated fixed point has $\Omega_\phi=1$ and $w_{\phi}=w_{\rm eff}\simeq -0.997$.
The initial conditions are $\Omega_{\phi i}=10^{-20}$ (top), $\Omega_{\phi i}=10^{-21}$ (middle), $\Omega_{\phi i}=10^{-22}$ (bottm) and $\Omega_{ri}=  0.998$ at the initial redshift $\log_{10}(1+z)=6$.
It is clear that the middle plot features the desired series of radiation, matter and dark-energy dominated epochs, though at the price of a severe fine tuning of the initial conditions.
On the contrary, a similar evolution can be obtained by the model presented in this paper without tuning the initial conditions, see figure~\ref{evo}.
}
\label{dedom}
\end{center}
\end{figure}

If the initial condition for $\Omega_{\phi i}$ is changed to the substantially different value $\Omega_{\phi i}=y_i^2=10^{-3}=\Omega_{m i}$
--
at this very step one usually encounters the fine-tuning problem as shown in figure~\ref{dedom}
--
one has basically the same trajectory, as shown in figure~\ref{evo} (first row, right panel). The end point is now at $\{ 0.0467, 0.834 \}\simeq \{ x_0, y_0\}$, showing how insensitive the evolution is with respect to the initial conditions. The same conclusion holds for any viable values of $\lambda$ and $\ppar$~\cite{paper2}.

The second row of figure~\ref{evo} shows the redshift evolution of the density parameters, and that this model achieves the matter dominated epoch after the radiation dominated one.
After the matter dominated era, the dark energy dominated era sets in as the scaling attractor with $\Omega_{m0}\simeq 0.3$ and $\Omega_{\phi 0}\simeq 0.7$ is reached.
The third row of figure~\ref{evo} shows the redshift evolution of the dark energy equation of state and of the effective equation of state.
Also shown is the effective equation of state for a flat $\Lambda$CDM model with the same present-day values of the density parameters.
The comparison of the plots of the left column with those of the right column shows that this model does not suffer from the fine tuning of initial conditions: the dynamics is basically unaltered -- see for example the evolution of the effective equation of state -- if the initial dark energy density is changed by seven orders of magnitude.

The dynamics shown in the second and third rows of figure~\ref{evo} is qualitatively similar to the standard coupled dark energy evolution if the parameters are chosen in order to approach an accelerated fixed point with $\Omega_\phi=1$ and $w_{\phi}=w_{\rm eff}=-1+ \frac{\lambda^2}{3}$.
However, the latter model suffers severely from the coincidence problem and its related fine tuning of initial conditions.
For the $\Lambda$CDM model and standard coupled dark energy models to work, one has to set the initial density parameter precisely to a value of about $\sim 10^{-20}$, tens of orders of magnitude smaller than the initial energy density of matter and radiation.
This is shown in figure~\ref{dedom} where the evolution of the density parameters for $ \{ \lambda , \ppar, n \}= \{0.1 ,0.3, 0 \}$ is shown for different initial values of the dark energy density.

Finally, the bottom row of figure~\ref{evo} shows the redshift evolution of the growth rate $f_g= \delta_m'/ \delta_m$ compared to the growth rate $f_{g,\Lambda}$ relative to the corresponding flat $\Lambda$CDM model defined above.
As it can be seen, the overall redshift dependence of the growth rate is qualitatively similar to the $\Lambda$CDM growth rate, even though moderately suppressed due to the fact that -- for this specific choice of parameters -- dark energy was subdominant but not negligible in the past.
Non negligible dark energy in the past could affect the CMB as Early Dark Energy studies have shown~\cite{Ade:2015rim}.

\section{Conclusions}
\label{sec:conclu}

We have discussed the possibility that dark energy could effectively couple to matter inhomogeneities.
Although only phenomenological, this model is interesting as it is possible to obtain a viable cosmology free of the coincidence problem for a wide range of parameters.
In particular, the current density of dark energy is given by the value of the coupling parameters $\lambda$ and $\ppar$ and is insensitive to the precise value of the initial conditions for the scalar field.

We proposed that this phenomenological interaction is due to the coupling of the scalar field to the vector perturbations associated with vorticity. Vorticity is indeed produced at very non-linear stages of the dark matter collapse when the single-stream irrotational description of the dark matter fluid ceases to be valid. Therefore, vorticity is highly correlated with matter perturbations, as also seen in dedicated numerical simulations which could be used to precisely evaluate the growth of the average vorticity in the universe.
Also left to future investigation is a comparison with present and future cosmological data in order to see if the model is viable and, in case, if upcoming observations could rule it out.

\noindent\paragraph{\bf\emph{Acknowledgements}} 
It is a pleasure to thank Luca Amendola, Celia Escamilla-Rivera, Oliver F.\ Piattella and Davi C.\ Rodrigues for useful comments and discussions.
VM is supported by the Brazilian research agency CNPq.

\bibliographystyle{elsarticle-num}
\bibliography{cosmo-lensing}

\end{document}